\documentclass[aps,prb,twocolumn,floats,preprintnumbers,superscriptaddress,amsmath,amssymb]{revtex4}
\usepackage{amsmath}
\usepackage{amssymb}
\usepackage{graphicx}

\begin{document}

\title{Oxygen adatoms at SrTiO$_{3}$(001): A density-functional theory study} 

%
\title{Oxygen adatoms at SrTiO$_{3}$(001): A density-functional theory study} 

\author{Hannes Guhl}
\affiliation{Fritz-Haber-Institut der Max-Planck-Gesellschaft, 
Faradayweg 4-6, D-14195 Berlin, Germany}
\affiliation{Leibniz Institut f\"ur Kristallz\"uchtung, Max-Born-Str. 2, D-12489 Berlin, Germany}

\author{Wolfram Miller}
\affiliation{Institut f\"ur Kristallz\"uchtung, Max-Born-Str. 2, D-12489 Berlin, Germany}

\author{Karsten Reuter}
\affiliation{Fritz-Haber-Institut der Max-Planck-Gesellschaft, 
Faradayweg 4-6, D-14195 Berlin, Germany}
\preprint{(submitted to Surf. Sci.)}
\received{July 15, 2009}

\begin{abstract}
We present a density-functional theory study addressing the 
energetics and electronic structure properties of isolated 
oxygen adatoms at the SrTiO$_3$(001) surface. 
Together with a surface lattice oxygen atom, the adsorbate 
is found to form a peroxide-type molecular species. 
This gives rise to a non-trivial topology of the potential energy 
surface for lateral adatom motion, with the most stable adsorption
 site not corresponding to the one expected from a continuation of 
the perovskite lattice. With computed modest diffusion barriers
 below 1\,eV, it is rather the overall too weak binding at both
 regular SrTiO$_3$(001) terminations that could be a critical 
factor for oxide film growth applications.
\end{abstract}

\maketitle

\section{Introduction}

Apart from its use in photo-catalytic and sensing 
applications \cite{mavroides76,merkle08}, the SrTiO$_3$(001) surface is 
also receiving increasing attention as a suitable substrate material for 
thin film growth \cite{henrich94,matijasevic96}. 
For the latter context the numerous reported surface 
reconstructions \cite{liang94,naito94,szot99,kubo01,erdman02,vonk05,lanier07,herger07},
 partly in sensitive dependence of the applied annealing temperature, 
indicate a complex surface kinetics, which needs to be understood and 
controlled when aiming at growth experiments tailored to the atomic-scale. 
This view of the surface being far from equilibrium under the employed 
experimental conditions \cite{herger07,liborio05} is 
corroborated by first-principles thermodynamic 
calculations \cite{liborio05,johnston04,heifets07}, which identified the unreconstructed 
surface as the only equilibrium termination. In the resulting focus on the
 adsorption kinetics, a first important step is to establish detailed
 insight into the nature of the bond and the concomitant binding and
 diffusion properties of all surface species involved. Studying the 
interaction of oxygen with SrTiO$_3$(001) is in this respect a natural 
starting point, in light of the important role played by oxygen exposure 
for surface preparation and growth or gas-sensing applications. 
Seemingly simple at first glance, preceding experimental and theoretical
 work has already revealed intriguing aspects even of the initial
 surface intermediates. 
Not untypical for wide band-gap oxides 
\cite{henrich94,che83}, spectroscopic and kinetic investigations suggest
 the existence of a molecular peroxide ``O$_2^{-2}$'' species at the 
surface \cite{bermudez80,merkle02}. 
This is confirmed by recent density-functional theory (DFT) calculations
 by Piskunov {\em et al.} \cite{piskunov06}, which demonstrate that 
this peroxide species results at both regular SrTiO$_3$(001) terminations 
actually from the bonding of an O adatom to a surface O ion of the 
oxide lattice.

The formation of such a dioxygen moiety may readily give rise to a 
non-trivial topology of the potential energy surface (PES) for lateral 
diffusion of adsorbed O atoms across the surface, with direct consequences 
for the role of O adatoms in the growth kinetics. 
Aiming to elucidate the latter we therefore employ DFT to explicitly map 
this PES for isolated O adatoms at both regular SrTiO$_3$(001) terminations
 and discuss the implications of the obtained energetic data with respect to the 
presence and mobility of this surface species. The computed surface electronic 
structure indeed suggests that the complex formed between the O adatom and 
lattice oxide ion may be characterized as a molecular peroxide species. 
While this surface bond is at both terminations strong enough to shift the 
most stable adsorption site away from the one that would correspond to a 
continuation of the perovskite lattice, the absolute bond strength
is only about the same as in gas-phase O$_2$ and the diffusion barrier with 
0.8\,eV quite modest. 
This points to the overall stabilization, rather than the mobility of a 
corresponding O adatom species as critical factor for the growth 
kinetics at this surface.

\section{Theory}

\begin{figure}
\begin{center}
  \includegraphics{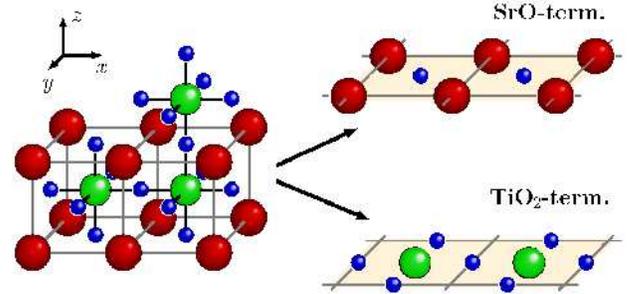}
\end{center}
\caption{\label{fig1}
(Color online) Schematic illustration of the crystal structure of cubic 
SrTiO$_3$ and its two regular (001) terminations, featuring a 
TiO$_2$ or SrO composition in the topmost layer. 
O atoms are depicted as small, light gray (blue) spheres, 
Ti atoms as medium-size, gray (green) spheres and Sr atoms as 
large, dark (red) spheres.}
\end{figure}

Identified as the only equilibrium surface structures 
\cite{liborio05,johnston04,heifets07}, the regular TiO$_2$- and SrO-terminations 
illustrated in Fig. \ref{fig1} form a natural basis for our fundamental study 
addressing the energetics and electronic structure of O adatoms at 
cubic SrTiO$_3$(001). 
All corresponding DFT calculations were performed with the CASTEP \cite{clark05} 
code using a plane-wave basis together with ultrasoft pseudopotentials \cite{vanderbilt90} 
as provided in the default library, and the GGA-PBE functional \cite{perdew96} to
treat electronic exchange and correlation (xc). Tests using the LDA did not yield any
qualitative differences with respect to the conclusions put forward below.
The surfaces were modeled with supercell geometries, using inversion-symmetric slabs 
based on the GGA-PBE optimized bulk lattice constant ($a_{0} = 3.938$\,{\AA}), 
with O adsorption at both sides and with a vacuum separation exceeding 13\,{\AA}. 
Apart from the central three layers of the slab, all substrate atomic positions 
were fully relaxed until the absoulute value of all ionic forces
dropped below 0.1\,eV/{\AA}, with further tightening of this criterion not yielding
any significant changes with respect to surface geometry and concomitant energetics.
This relaxation was equally applied to the $z$-height of the O adatom, with its 
$x$ and $y$ coordinates frozen at different positions to map the PES 
for lateral adatom motion.

The centrally targeted energetic quantity is the binding energy of an O adsorbate,
defined as
\begin{equation}
\label{eq1}
E_{\rm b} \;=\; \frac{1}{2} \left[ E_{\rm O@surf} - E_{\rm surf} - E_{\rm O_2(gas)} \right] \quad ,
\end{equation}
where $E_{\rm O@surf}$ is the total energy of the O covered surface, 
$E_{\rm surf}$ the total energy of the clean surface, $E_{\rm O_2(gas)}$ 
the total energy of the gas-phase O$_2$ molecule (all three computed at 
the same plane-wave cutoff), and the factor $1/2$ accounts for the fact that 
adsorption is at both sides of the slab. The binding energy is thus referenced 
to a free O$_2$ molecule in its spin-triplet ground state (calculated in a 
$(18 \times 18 \times 18)$\,{\AA} supercell), and a negative value of 
$E_{\rm b}$ indicates that adsorption is exothermic. 
Both $E_{\rm O@surf}$  and $E_{\rm surf}$ were furthermore consistently 
computed in a non spin-polarized 
way, with systematic tests confirming that spin-polarization at the surface
does not play any significant role for the description of all relevant low-energy 
parts of the PES.

Systematic tests  with one O adatom at different high-symmetry 
sites in a $(1 \times 1)$  surface unit-cell verified that 
 $E_{\rm b}$  is numerically converged with respect to the  basis set to 
within $\pm 20$\,meV when employing a plane wave kinetic energy cutoff 
of 430\,eV.
Due to long-range geometric relaxations \cite{carrasco06} achieving 
a similar convergence with respect to the number of surface unit cells in 
the slab model is more challenging.
This concerns prominently those adsorption sites where
the O adatom protrudes significantly from the surface, e.g. on-top of the 
lattice Ti-atom on the TiO$_2$ termination.
However, at the same time such sites are energetically so unfavorable 
that they do not even belong 
to the saddle point regions of the adatom PES that are relevant for the diffusion
barriers. 
Consequently, we  employed $(3 \times 3)$ surface unit-cells with a
$(2 \times 2 \times 1)$ Monkhorst-Pack grid \cite{monkhorst99} 
for the reciprocal space integrations.
Test calculations in 
 larger surface unit-cells indicate that this set-up provides a 
 very good description 
 of the isolated adatom limit with respect to the most stable site and diffusion 
 barriers addressed here.
The slab models comprised in total  seven and 11  atomic layers 
for the TiO$_{2}$- and SrO-termination, respectively, 
 which ensure that all energetically relevant parts of the adatom PES 
discussed below are converged to at least within $\pm 50$\,meV. 

\begin{figure*}
\begin{center}
  \includegraphics[angle=0]{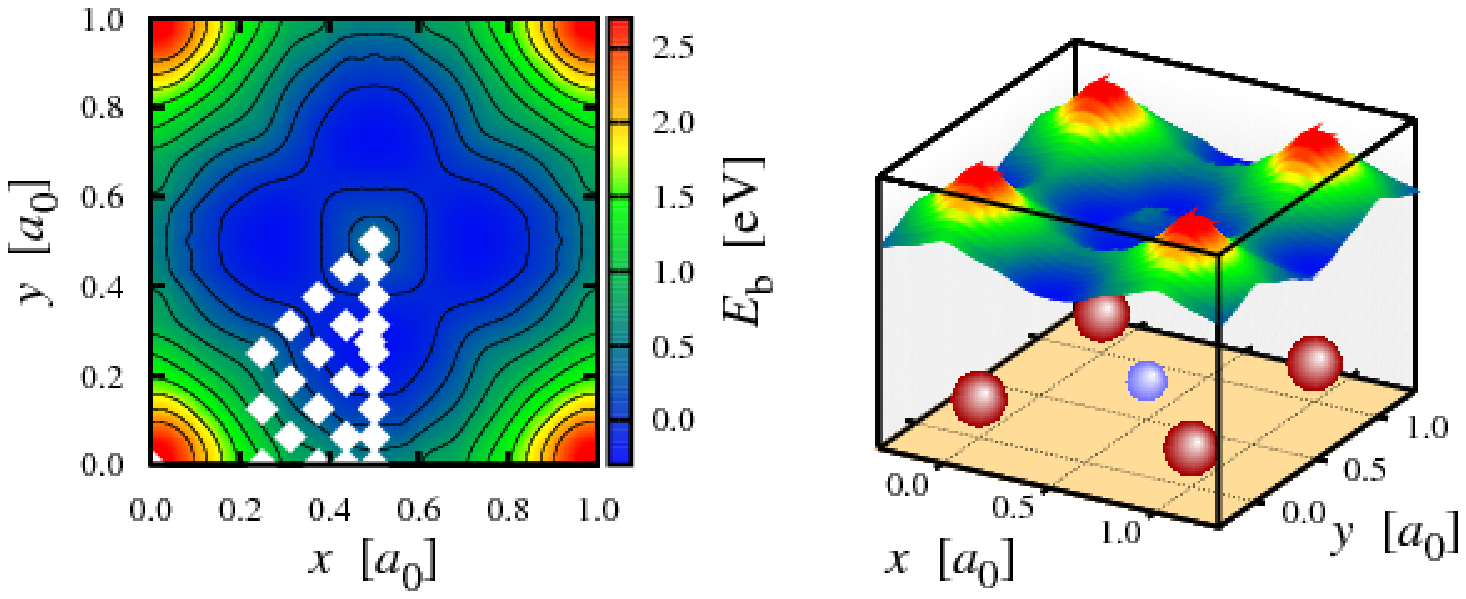}
  \includegraphics[angle=0]{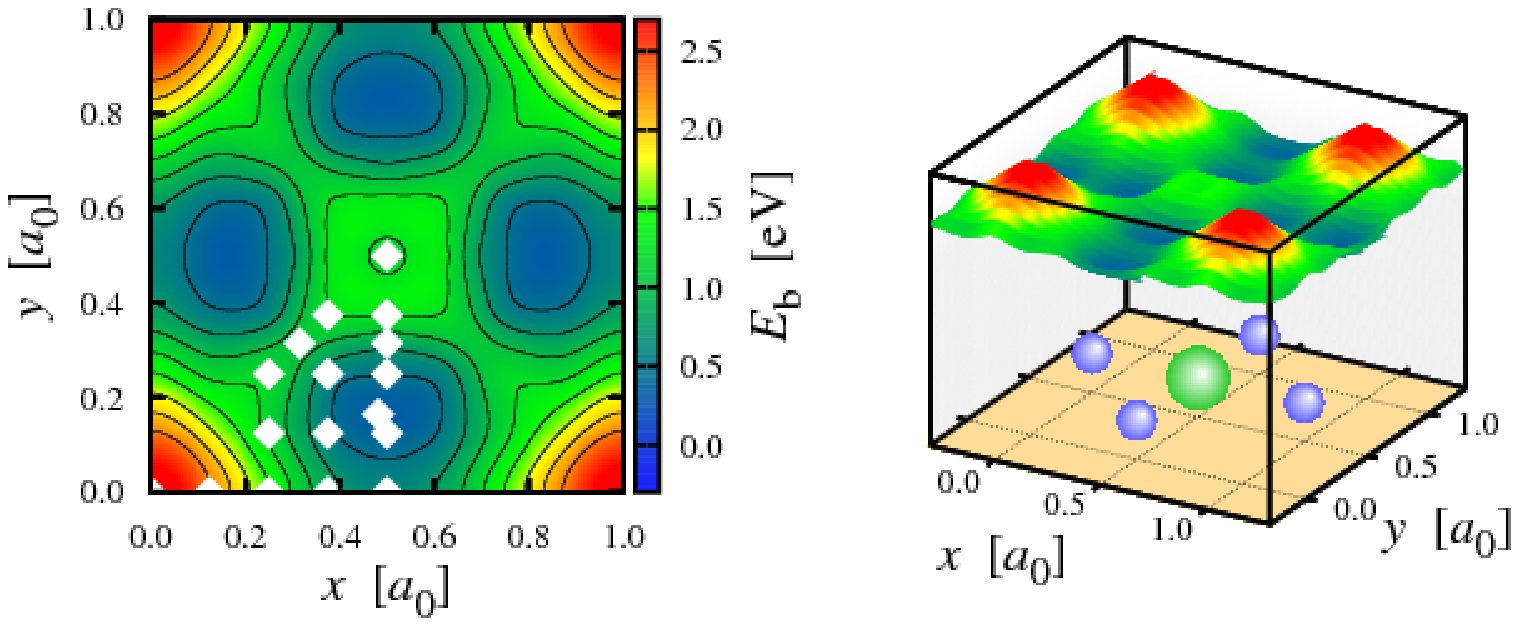}
\end{center}
\caption{\label{fig2}
(Color online) Potential energy surface PES($x$,$y$) for lateral motion
of an O adatom at the SrO- and TiO$_2$-termination, upper and lower panels 
respectively. The lateral coordinates $x$ and $y$ are given in units of the 
bulk lattice constant $a_{0} = 3.938$\,{\AA}, with the shown range for each
 termination exactly corresponding to the surface unit-cell definition shown 
in Fig. \ref{fig1}. At the SrO-termination, the position (0,0) is thus directly 
atop a surface Sr atom, while at the TiO$_2$-termination the position 
(1/2, 1/2) is atop a surface Ti atom. 
The white diamonds denote the actually 
calculated points within the irreducible wedge and neighboring isolines 
correspond to a difference of 0.25 eV.
}
\end{figure*}

\section{Results}

\subsection{Geometric structure and energetics}

\begin{figure}
\begin{center}
  \includegraphics{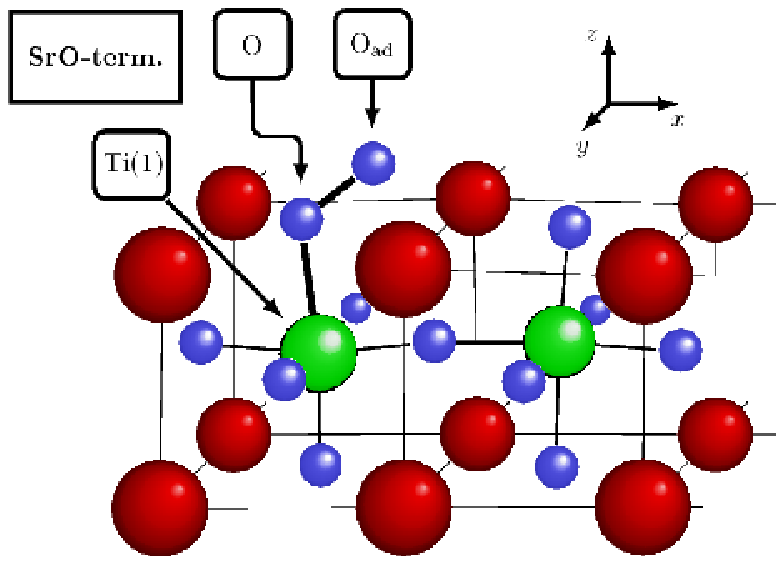}
  \includegraphics{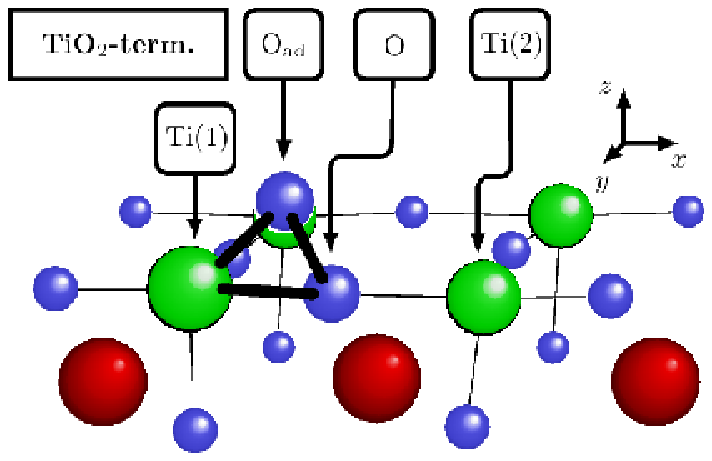}
\end{center}
\caption{\label{fig3}
(Color online) Perspective view of the most stable O adsorption geometry 
at the SrO- and TiO$_2$-termination, upper and lower panel respectively. 
Atoms specifically referred to in the text are labeled. O atoms are depicted 
as small, light gray (blue) spheres, Ti atoms as medium-size, gray (green)
 spheres and Sr atoms as large, dark (red) spheres.}
\end{figure}

Figure \ref{fig2} summarizes the computed PES for lateral O adatom motion 
at both regular terminations. As suspected, the shape is in both cases 
non-trivial, with the global minimum, i.e. most stable adsorption site,
 not corresponding to the one expected from a mere continuation of the 
perovskite lattice.
 The geometric adsorption structure at this most stable 
site shown in Fig. \ref{fig3} for both terminations is indeed very consistent 
with the anticipated reason \cite{piskunov06} for this PES topology in form 
of strong bonding between the adsorbate and a SrTiO$_3$(001) surface oxygen 
anion ({\em vide infra}). 
At the SrO-termination this directly coordinated lattice oxygen atom is 
pulled out of the surface while the underlying Ti atom, denoted as 
Ti(1) in Fig. \ref{fig3}, is pressed into the bulk material,
thereby stretching  the Ti-O bond from  
1.97\,{\AA} to 2.20\,{\AA}. 

On the other hand the distance to the adsorbed O$_{\rm ad}$ atom is quite 
short, and with $1.52$\,{\AA} rather characteristic for a molecular 
peroxide ``O$_2^{-2}$'' species \cite{henrich94,che83}. 
With respect to the surface normal the dioxygen complex is tilted with 
a Ti(1)-O-O$_{\rm ad}$ angle of $120^\circ$ and points between two surface 
Sr atoms. 
However, the barrier for a $90^\circ$ rotation of the complex around the 
$z$-axis to point between an adjacent pair of Sr atoms is only 0.24\,eV, 
cf. Fig. \ref{fig2}. One could thus view the dioxygen moiety as a tilted, 
singly-bonded $\eta^1$ ``end-on'' configuration with respect to the 
underlying Ti(1) cation in the second layer, with the tilt direction 
exhibiting some dynamics at finite temperature.

At the TiO$_2$-termination only atoms in the topmost layer are 
significantly affected by the presence of the oxygen adatom. 
The oxygen-oxygen bond measures 1.47\,{\AA} and is therewith again in 
the ballpark expected for a molecular peroxide species. 
It is tilted towards one of the adjacent Ti surface atoms denoted as 
Ti(1) in Fig. \ref{fig3}, such that the distance between O$_{\rm ad}$ 
and the latter is 1.91\,{\AA} and the Ti(1)-O-O$_{\rm ad}$ tilt angle 
is $62^\circ$. 
This geometry is thus more reminiscent of an asymmetric, doubly-bonded 
$\eta^{2}$ ``side-on'' configuration with respect to the coordinated 
Ti(1) cation, which is in fact the more usual form in dioxygen-Ti 
complexes \cite{vaska76} and with the respective bond-lengths and angles 
of the Ti(1)-O-O$_{\rm ad}$ moiety agreeing again specifically with those 
of peroxotitanium complexes \cite{schwarzenbach70}. 
Despite this coordination a flipping motion of the dioxygen moiety to 
the adjacent Ti atom denoted as Ti(2) in Fig. \ref{fig3} has only a 
small computed barrier of 0.30\,eV as shown in Fig. \ref{fig2}, which 
again indicates some dangling dynamics at finite temperature.

Even in this most stable site at the TiO$_2$-termination the computed 
binding energy is with 0.30\,eV still endothermic with respect to 
gas-phase O$_2$. However, due to the known shortcomings of the employed 
gradient-corrected xc functional such absolute binding energy values have 
to be considered with care. In this respect, it is quite gratifying that 
the preceding study by Piskunov and coworkers \cite{piskunov06}, which 
employed a hybrid functional including some non-local Hartree-Fock exchange, 
reports a rather similar binding energy of 0.66\,eV at the same most 
stable site and with nearly identical geometry parameters. 
In fact, when redoing our calculations in the smaller $(2 \times 2)$ 
surface unit-cell employed by them we arrive at a virtually coinciding 
$E_{\rm b} = 0.60$\,eV.
This seems to indicate that a proper account 
of the long-range geometric relaxations enabled in the larger surface 
unit-cells accessible with the computationally much less demanding 
semi-local functional have a larger effect on the binding energetics 
than the improved xc description introduced by the hybrid functional. 
Furthermore, we believe that the additional GGA advantage of being able 
to perform a much more thorough exploration of the PES has direct 
consequences for the comparison of the results obtained for the 
SrO-termination. 
Within the understanding of the PES topology shown in 
Fig. \ref{fig2} it appears as if the restricted testing of three possible 
high-symmetry adsorption sites has misled the authors of 
ref. \cite{piskunov06} to erroneously identify the transition-state 
for the rotational motion of the oxygen moiety around the $z$-axis as 
most stable adsorption site, while the PES area around the above described 
most stable site of lower symmetry was not scanned in their study. 
Recomputing this transition state in the smaller surface unit-cell we 
again achieve excellent agreement between our GGA-PBE and their reported 
hybrid functional energetics, which fortifies our confidence in the here 
calculated slightly exothermic $E_{\rm b} = -0.20$\,eV at the most stable 
site shown in Fig. \ref{fig3}.

At least at the TiO$_2$-termination the binding is thus 
clearly too weak to generate O adatoms through dissociative 
adsorption at the ideal terraces. 
This is commonly expected for 
insulating oxides \cite{henrich94} and has been verified explicitly for 
related materials like SnO$_2$ \cite{oviedo01}, MgO \cite{kantorovich97} 
and TiO$_2$ \cite{rasmussen04}. Considering the high vacancy formation 
energies \cite{carrasco06}, it would, however, energetically clearly be 
possible to create an O adatom by adsorbing an oxygen molecule into a 
surface O vacancy at both regular terminations. 
Once established at the 
surface, the computed barriers for diffusive motion to the adjacent 
binding site of 0.81\,eV and 0.68\,eV at the SrO- and TiO$_2$-termination 
respectively, cf. Fig. \ref{fig2}, demonstrate that the O-O$_{\rm ad}$ 
bond is then much more readily broken
compared to the desorption of the O-O$_{\rm
  ad}$ molecule
 so that O adatoms should be a 
relatively mobile species.

\subsection{Electronic structure}

\begin{figure*}
\begin{center}
  \includegraphics[angle=0]{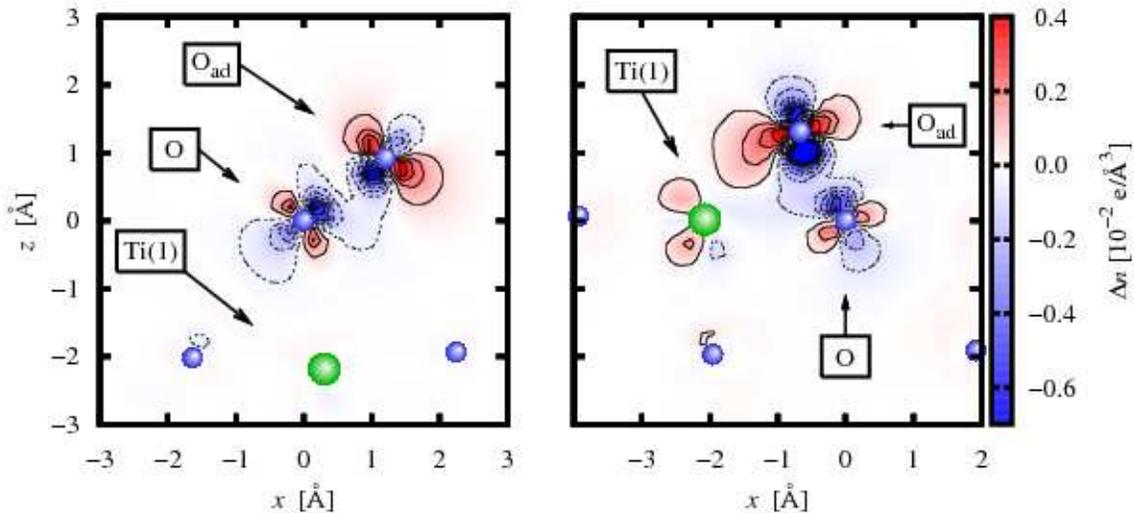}
\end{center}
\caption{\label{fig4}(Color online) Difference electron 
  density $\Delta n({\bf r})$ plot highlighting 
the formation of a dioxygen moiety at the SrO- and TiO$_2$-termination, 
left and right panel respectively. 
Shown is the charge rearrangement within the plane perpendicular to the 
surface defined by the O$_{\rm ad}$ adatom and its directly coordinated lattice 
O and Ti(1) partners, cf. Fig. \ref{fig3}, with red (blue) 
areas with solid (dashed) contour lines depicting charge depletion and accumulation, respectively. 
Neighboring isolines indicate a difference of $10^{-3} e/{\text{\AA}^3}$.
The superimposed spheres indicate the position of atoms within the plane, using the color 
and size scheme for the individual species as in Figs.~\ref{fig1} and \ref{fig3}.
}
\end{figure*}

The preceding section has discussed the geometric structure and adsorption 
energetics within the picture of a quasi-molecular species formed between 
the O adatom and a directly coordinated lattice O anion. 
This view is nicely supported by the adsorbate-induced change of the 
electron density shown in Fig. \ref{fig4}. This quantity, commonly referred 
to as ``difference electron density'' \cite{scheffler00}, is obtained by 
subtracting from the electron density of the O@SrTiO$_3$(001) system both 
the electron density of the clean SrTiO$_3$(001) surface and that of an 
isolated oxygen atom. Here, the atomic positions of SrTiO$_3$(001) and of 
the O atom are taken to be same as those of the relaxed adsorbate system. 
In this way, the presentation highlights the charge rearrangement upon 
bond formation, which as clear from Fig. \ref{fig4} primarily takes place between 
the O adatom and its lattice O partner. 

\begin{figure}
\begin{center}
  \includegraphics{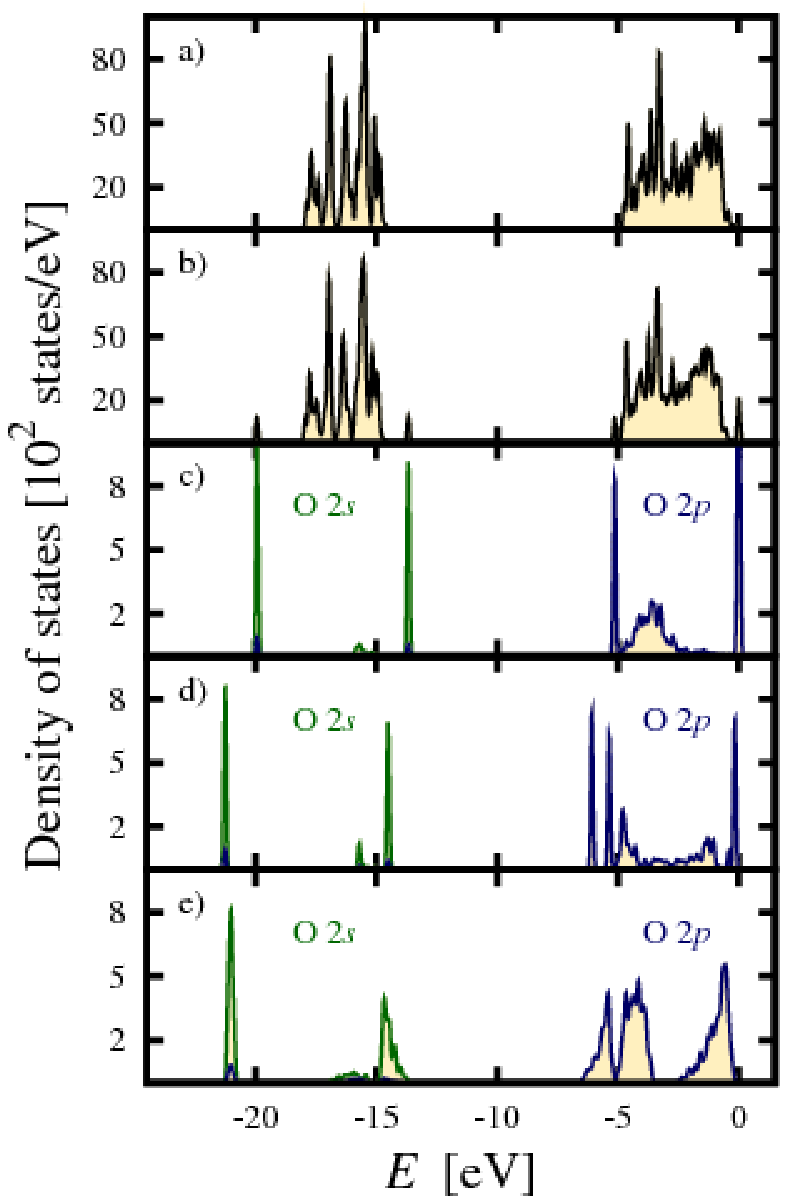}
\end{center}
\caption{\label{fig5}
(Color online) Density of states (DOS) analysis supporting the interpretation 
of the formed O$_{\rm ad}$-O complex in terms of a molecular peroxide species. 
The topmost panels show the total DOS of the topmost two surface layers of the 
SrO-termination of clean SrTiO$_3$(001) before (a) and after (b) O  adsorption. 
The next two panels show the DOS projected onto the $2s$ (green line) and $2p$ (blue line) 
states of the O$_{\rm ad}$ atom and its coordinated lattice O partner for the SrO- (c) 
and the TiO$_2$-termination (d), identifying in particular two sharp states above and 
below the O $2s$ group as fingerprints for the formed O$_{\rm ad}$-O moiety. 
Panel (e) shows the same projected DOS of bulk SrO$_2$ exhibiting essentially the 
same  characteristics. The bulk valence band maximum (as approximately reached in 
the central slab layer) is taken as zero reference in all panels.}
\end{figure}

Some further quantification of this adsorbate-induced charge redistribution 
can be obtained from an inspection of the Mulliken \cite{mulliken55} and 
Hirshfeld \cite{hirshfeld77} charges on the various surface species. 
With the two charge partitioning schemes yielding the usual differences 
in absolute numbers, they nevertheless unanimously predict the same trend 
in that the charge surplus at the lattice O anion is essentially evenly 
distributed over the formed dioxygen moiety after adsorption. 
In other  words, while the charges on all other atoms in the 
topmost two layers are virtually unaffected by the adsorbed O atom, 
the charge on the directly  coordinated lattice O anion gets about 
halved with the O adatom exhibiting a charge of roughly equal magnitude. 
For the Mulliken (Hirshfeld) scheme 
this means specifically that at the TiO$_2$-termination the computed charge 
of $-0.73e$ ($-0.33e$) at the surface O anion, diminishes to $-0.49e$ ($-0.16e$) 
upon adsorption with the O adatom exhibiting a similar charge of $-0.38e$ 
($-0.16e$). 
At the SrO-termination the equivalent numbers are $-0.83e$ ($-0.38e$) 
and $-0.54e$ ($-0.20e$) for the surface anion before and after O adsorption, 
as well as $-0.58e$ ($-0.30e$) for the O adatom. 
The thereby suggested formation of a molecular peroxide ``O$_2^{-2}$'' 
species is corroborated by an analysis of the density of states (DOS) at 
the surface, which we focus for the present purposes on the O $2s$ dominated 
states around 20\,eV below the bulk valence band maximum.
Figure \ref{fig5} compiles the corresponding calculations and compares 
them to the  oxygen-projected DOS of bulk SrO$_2$, a material 
in which O$_2^{-2}$-type  bonding occurs naturally. 
Apparently, O adsorption at SrTiO$_3$(001) leads 
at both terminations to two sharp fingerprint states above and below the 
regular SrTiO$_3$ O $2s$ group, which almost perfectly match the energetic 
position of the equivalent states in bulk SrO$_2$. 
This confirms the 
interpretation in terms of of a molecular peroxide species, the bond of 
which leads to the non-trivial PES topology for O adatom diffusion described 
in Section 3.A.

\section{Conclusions}

In conclusion we have presented a density-functional theory study analyzing 
the potential-energy surface for lateral diffusion of O adatoms at both
 regular terminations of SrTiO$_3$(001). 
The O adatom is found to form a 
quasi-molecular ``O$_2^{-2}$''-type species with a directly coordinated 
surface O anion of the substrate lattice. 
This bond is strong enough to 
shift the most stable adsorption site away from the one expected from a 
mere continuation of the perovskite lattice, and gives rise to a non-trivial 
PES topology that allows for some dangling dynamics of the dioxygen moiety at finite 
temperature. 
In addition, the computed barriers for diffusive hops to 
adjacent binding sites are at both terminations well below 1\,eV, so that 
mobility of O adatoms is not expected to represent a bottleneck at typical 
film growth conditions.
On the other hand, the overall bond strength is at 
both terminations only about equal to the one of a O$_2$ molecule, and 
specifically slightly exothermic at the SrO-termination and slightly 
endothermic at the TiO$_2$-termination with respect to this gas-phase 
reference. 
At least at the latter termination the binding is thus clearly 
too weak to generate O adatoms through dissociative adsorption at ideal 
SrTiO$_3$(001) terraces. 
In view of the high O surface vacancy formation 
energies, an at least energetically much more plausible alternative would 
be that O adatoms are created by adsorbing an oxygen molecule into such a 
defect. Once established at the surface, the computed modest diffusion 
barriers show that the ``O$_2^{-2}$'' peroxide-type bond is broken rather 
readily, suggesting that such O adatoms would provide a mobile species 
that could anneal further defects or attach to created islands.

\section*{Acknowledgements}

All calculations have been performed on the SGI Altix ICE- and XE- 
computing clusters of the North-German Supercomputing Alliance (HLRN) 
hosted at  Konrad-Zuse-Zentrum f\"{u}r Informationstechnik in 
Berlin (ZIB) and at  Regionales Rechenzentrum f\"{u}r 
Niedersachsen (RRZN) at Leibniz Universtit\"{a}t  Hannover. 
We are particularly grateful 
to B.~Kallies for technical support.


\begin{thebibliography}{99}

\bibitem{mavroides76}
J.G. Mavroides, J.A. Kafalas, and D.F. Kolesar, Appl. Phys. Lett. {\bf 28}, 241 (1976).

\bibitem{merkle08}
R. Merkle and J. Maier, Angew. Chem. Int. Ed. {\bf 47}, 3874 (2008).

\bibitem{henrich94}
V.E. Henrich and P.A. Cox, {\em The Surface Science of Metal Oxides},
Cambridge University Press, Cambridge (1994)

\bibitem{matijasevic96}
V.C. Matijasevic, B. Ilge, B. St\"auble-P\"umpin, G. Rietveld,
F. Tuinstra, and J.E. Mooij, Phys. Rev. Lett. {\bf 76}, 4765 (1996).

\bibitem{liang94}
Y. Liang and D.A. Bonnell, Surf. Sci. {\bf 310}, 128 (1994).

\bibitem{naito94}
M. Naito and H. Sato, Physica C {\bf 229}, 1 (1994).

\bibitem{szot99}
K. Szot and W. Speier, Phys. Rev. B {\bf 60}, 5909 (1999).

\bibitem{kubo01}
T. Kubo and H. Nozoye, Phys. Rev. Lett. {\bf 86}, 1801 (2001).

\bibitem{erdman02}
N. Erdman, K.R. Poeppelmeier, M. Asta, O. Warschkow, L.D. Marks, and D.E. Ellis,
Nature {\bf 419}, 55 (2002).

\bibitem{vonk05}
V. Vonk, S. Konings, G. van Hummel, S. Harkema, and H. Graafsma,
Surf. Sci. {\bf 595}, 183 (2005).

\bibitem{lanier07}
C.H. Lanier, A. van~de Walle, N. Erdman, E. Landree, O. Warschkow, A. Kazimirov, K.R. Poeppelmeier,
J. Zegenhagen, M. Asta, and L.D. Marks, Phys. Rev. B {\bf 76}, 045421 (2007).

\bibitem{herger07}
R. Herger, P.R. Willmott, O. Bunk, C.M. Schleputz, B.D. Patterson, B. Delley,
V.L. Shneerson, P.F. Lyman, and D.K. Saldin, Phys. Rev. B {\bf 76}, 195435 (2007).

\bibitem{liborio05}
L.M. Liborio, C.G. Sanchez, A.T. Paxton, and M.W. Finnis,
J. Phys.: Condens. Matter {\bf 17}, L223 (2005).

\bibitem{johnston04}
K. Johnston, M.R. Castell, A.T. Paxton, and M.W. Finnis,
Phys. Rev. B {\bf 70}, 085415 (2004).

\bibitem{heifets07}
E. Heifets, S. Piskunov, E.A. Kotomin, Y.F. Zhukovskii, and D.E. Ellis,
Phys. Rev. B {\bf 75}, 115417 (2007).

\bibitem{che83}
M. Che and A.J. Trench, Adv. Catal. {\bf 32}, 1 (1983).

\bibitem{bermudez80}
V.M. Bermudez and V.H. Ritz, Chem. Phys. Lett. {\bf 73}, 160 (1980).

\bibitem{merkle02}
R. Merkle and J. Maier, Phys. Chem. Chem. Phys. {\bf 4}, 4140 (2002).

\bibitem{piskunov06}
S. Piskunov, Y.F. Zhukovskii, E.A. Kotomin, E. Heifets, and D.E. Ellis,
Proc. of Mat. Res. Soc. Symposia {\bf 894}, 295 (2006).

\bibitem{clark05}
S. Clark, M.D. Segall, C.J. Pickard, P.J. Hasnip, M.I.J. Probert, K. Refson, and M.C. Payne, 
Z. Kristallogr. \textbf{220}, 567 (2005).

\bibitem{vanderbilt90}
D. Vanderbilt, Phys. Rev. B \textbf{41}, 7892 (1990).

\bibitem{perdew96}
J.P. Perdew, K. Burke, and M. Ernzerhof, Phys. Rev. Lett. {\bf 77}, 3865 (1996).

\bibitem{monkhorst99}
H. Monkhorst and J. Pack, Phys. Rev. B \textbf{13}, 5188 (1976).

\bibitem{carrasco06}
J. Carrasco, F. Illas, N. Lopez, E.A. Kotomin, Y.F. Zhukovskii, R.A. Evarestov, Y.A. Mastrikov, S. Piskunov,
and J. Maier, Phys. Rev. B {\bf 73}, 064106 (2006).

\bibitem{vaska76}
L. Vaska, Acc. Chem. Res. {\bf 9}, 175 (1976).

\bibitem{schwarzenbach70}
D. Schwarzenbach, Inorg. Chem. {\bf 9}, 2391 (1970).

\bibitem{oviedo01}
J. Oviedo and M.J. Gillan, Surf. Sci. {\bf 490}, 221 (2001).

\bibitem{kantorovich97}
L.N. Kantorovich and M.J. Gillan,  Surf. Sci. {\bf 374}, 373 (1997).

\bibitem{rasmussen04}
M.D. Rasmussen, L.M. Molina, and B. Hammer, J. Chem. Phys. {\bf 120}, 988 (2004).

\bibitem{scheffler00}
M. Scheffler and C. Stampfl, {\em Theory of Adsorption on Metal Substrates}, 
In: Handbook of Surface Science, Vol. {\bf 2}: Electronic Structure, 
(Eds.) K. Horn and M. Scheffler, Elsevier, Amsterdam (2000).

\bibitem{mulliken55}
R.S. Mulliken, J. Chem. Phys. {\bf 23}, 1833 (1955).

\bibitem{hirshfeld77}
F.L. Hirshfeld, Theor. Chim. Acta {\bf 44}, 129 (1977).

\end{thebibliography}
\end{document}